\documentclass[aps,prc,showpacs,superscriptaddress,twocolumn]{revtex4}
 \usepackage{graphics}
 \usepackage{amsmath,amssymb}
 \usepackage{epsfig}
 \usepackage[english]{babel}

\begin{document}

\title{Reply to Comment on ``Multiple locations of boron atoms in the exohedral and endohedral C$_{60}$ fullerene" by J. Xu and G.-L. Hou}

\author{A. V. Bibikov}

\affiliation{Skobeltsyn Institute of Nuclear Physics Lomonosov
Moscow State University, 119991 Moscow, Russia}

\affiliation{National Research Nuclear University MEPhI, 115409,
Kashirskoe shosse 31, Moscow, Russia}

\author{A. V. Nikolaev}

\affiliation{Skobeltsyn Institute of Nuclear Physics Lomonosov
Moscow State University, 119991 Moscow, Russia}

\affiliation{National Research Nuclear University MEPhI, 115409,
Kashirskoe shosse 31, Moscow, Russia}

\affiliation{School of Electronics, Photonics and Molecular Physics, Moscow Institute of Physics and Technology, 141700, Dolgoprudny, Moscow region, Russia}

\author{I. V. Bodrenko}

\affiliation{CNR/IOM - Cagliari, Cittadella Universitaria di Monserrato, SP Monserrato-Sestu Km 0.700 I-09042 Monserrato, Italy}

\author{P. V. Borisyuk}

\affiliation{National Research Nuclear University MEPhI, 115409,
Kashirskoe shosse 31, Moscow, Russia}

\author{E.~V.~Tkalya}

\affiliation{P. N. Lebedev Physical Institute of the Russian
Academy of Sciences, 119991, 53 Leninskiy pr., Moscow, Russia}

\affiliation{National Research Nuclear University MEPhI, 115409,
Kashirskoe shosse 31, Moscow, Russia}

\affiliation{Nuclear Safety Institute of RAS, Bol'shaya Tulskaya
52, Moscow 115191, Russia}


\date{\today}

\begin{abstract}
In three out of five cases considered in our work, DFT calculations presented by Xu and Hou in their Comment give the same ground state confirmations.
On the other hand, depending on the choice of the exchange-correlation functional, the geometry optimization within DFT results in different ground state
confirmations for B@C$_{60}$ and BC$_{60}$, Table I of the Comment. Therefore, the energy balance between
nearest confirmations in these molecular complexes is subtle, and various methods can give different ground state structures.
Consequently, the results of our method -- the Hartree-Fock (HF) approach with the second order M{\o}ller-Plesset perturbation theory (MP2) --
should be compared with the DFT results on equal ground, we cannot agree that the DFT method used in the Comment is superior to HF-MP2.
In the Reply, we also present additional HF calculations with the 6-31G* basis set (used in the Comment for the geometry optimization) to show that
the polarization functions do not change the ground state confirmations obtained by us earlier at the HF/6-31G level.
\end{abstract}


\maketitle


In their Comment on our work \cite{Bib}, Xu and Hou \cite{Xu} claim that we use the method (Hartree-Fock with the second order M{\o}ller-Plesset perturbation theory MP2,
HF-MP2) which is not ``sufficiently reliable'',
while our results could be incorrect.
Xu and Hou put forward their density functional theory (DFT) calculations.
In our reply to their criticism of our work, we discuss the fact that in these molecular complexes different methods give different ground state confirmations,
analyse in detail equilibrium geometries
found in our study \cite{Bib} and the Comment \cite{Xu}
and investigate the role of the 6-31G* basis set. For that, using the GAMESS package \cite{GAMESS} we have carried out additional HF-calculations
at distinct equilibrium geometries of B$_2$@C$_{60}$ and B@C$_{60}$B obtained in Refs.~\cite{Bib} (HF-MP2) and the Comment \cite{Xu} (DFT), Tables \ref{tab1} and \ref{tab2}.
%
\begin{table}
\caption{
Comparison of the HF energies of the ground state  of B$_2$@C$_{60}$ for two distinct equilibrium geometries
found in Ref.~\cite{Bib} ($E_1$, in a.u. with $S=2$) and the Comment~\cite{Xu} ($E_2$, in a.u. with $S=0$),
$\triangle E=E_1(S=2) - E_2(S=0)$, in eV.
\label{tab1} }
\begin{ruledtabular}
\begin{tabular}{l  c  c  c  c }
 method/basis  & $E_1(S=2)$, a.u. &  $E_2(S=0)$, a.u. & $\triangle E$, eV \\
\tableline
  HF/6-31G           &  -2320.06230    &  -2319.95969        & -2.792 \\
  HF/6-31G$^{*}$     &  -2320.92059    &  -2320.87520        & -1.235 \\
  RIMP2/6-31G        &  -2325.51493    &  -2325.49605        & -0.514 \\

\end{tabular}
\end{ruledtabular}
\end{table}
%
\begin{table}
\caption{
Comparison of the HF energies of the ground state of B@C$_{60}$B for two distinct equilibrium geometries
found in Ref.~\cite{Bib} ($E_1$, in a.u.) and the Comment~\cite{Xu} ($E_2$, in a.u.),
$\triangle E=E_1 - E_2$, in eV. Spin quantum number $S=0$.
\label{tab2} }
\begin{ruledtabular}
\begin{tabular}{l  c  c  c  c }
 method/basis  & $E_1$, a.u. &  $E_2$, a.u. & $\triangle E$, eV \\
\tableline
  HF/6-31G           &  -2320.01492    &  -2319.97483        & -1.091 \\
  HF/6-31G$^{*}$     &  -2320.88715    &  -2320.86574        & -0.583 \\
  RIMP2/6-31G        &  -2325.49143    &  -2325.51117        & +0.537 \\

\end{tabular}
\end{ruledtabular}
\end{table}

First of all, we note that although the DFT approach is a powerful tool in modern quantum chemistry
and sometimes can give very good results, a perfect comparison with experimental data and other {\it ab initio} methods is not guaranteed {\it a priori}.
The Kohn–Sham (KS) foundation of DFT is well established, and the ``method is capable, in
principle, of yielding exact results, but because the equations of the Kohn–Sham (KS)
method contain an unknown functional that must be approximated, the KS formulation
of DFT yields approximate results" (p.~555 of Ref.~\cite{QC}).
The HF method with the MP2 treatment, which we have used in our research, is
a choice of {\it ab initio} approach, which takes into account dynamical correlation effects.
HF-MP2 remains one of the major tools used in modern quantum chemistry and can hardly be called ``oversimplified computer methodology" \cite{Xu}.
In Refs.\ \cite{Mili,Sher} it is used for comparison with high level coupled cluster benchmark calculations of the benzene dimer.

In the present case, as can be seen from Table I of Ref.~\cite{Xu}, the situation is ambiguous even within the DFT method, where
various DFT exchange-correlation parts result in different ground state geometries of C$_{60}$ with a single boron atom.
The highest level of the geometry optimization in the Comment is realized in $\omega$B97XD/6-31G* (denoted in the following as DFT-I) and BPW91/6-31G* (denoted as DFT-II)
with the 6-31G* basis set.
For a single boron atom with C$_{60}$, Table I of Ref.~\cite{Xu}, we immediately see contradictory results for the ground state confirmations.
For the endohedral case, i.e. B@C$_{60}$, the configuration Iso1 lies at 0.25 eV {\it lower} than the configuration Iso2 in DFT-I,
whereas the same configuration lies at 0.84 eV {\it higher} than Iso2 in DFT-II. The same ambiguity applies to the exohedral BC$_{60}$: with DFT-I the configuration
Iso1 ($\eta^{2(6-6)}$) lies at 0.22 eV {\it lower} than Iso2 ($\eta^{2(5-6)}$), while with DFT-II Iso1 ($\eta^{2(6-6)}$) lies at 0.02 eV {\it higher}
than Iso2 ($\eta^{2(5-6)}$).
This ambiguity in DFT results is not discussed and explained in the Comment, and therefore we do not think that the high level single point DFT calculations
obtained at these geometries and listed in the last two lines in Table I of the Comment \cite{Xu}
should be considered as a benchmark of the calculations.
Our opinion is that different ground state confirmations obtained with different DFT exchange-correlation functionals, indicate that the energy balance between
two nearest confirmations is fragile, and it can be easily changed in various methods.
There are other examples of such a situation in the literature, for example, the optimal position of
the benzene dimer, which can be different in various methods \cite{Sher}.
This problem requires a careful study with high level methods, {\it a priory}
one cannot say that some methods are ``unreliable" and should be discarded.
Therefore, we believe that the results of DFT should be compared with ours on equal ground.

We now discuss the results presented in the Comment.
Although the authors have not written it explicitly, their conclusions largely support our calculations.
For a single boron atom inside C$_{60}$ and outside C$_{60}$ the lowest conformations, Fig.~1 (upper left and middle panels) of the Comment \cite{Xu},
are in qualitative agreement with our findings (configurations (1,0) in Fig.~1 and (0,1) in Fig.~5(a,b,c) of Ref.~\cite{Bib}).
The same holds for the B$_2$ molecule outside the fullerene, B$_2$C$_{60}$, the configuration 0,2(mol) in Fig.~7(a,b) of \cite{Bib} and Iso1 in the upper right panel
of Fig.~1 of \cite{Xu}.
{\it Thus, in three out of five cases written in our abstract and conclusions \cite{Bib}, the molecular ground state structures found in HF-MP2 and DFT are the same.}
(The case of two carbon atoms in C$_{60}$ substituted with two boron atoms (the right lower panel in Fig.~1 of \cite{Xu}) has not been studied
thoroughly in \cite{Bib},
only few comparative calculations have been done {\it without mentioning this case in our abstract and conclusions}.)
Let us consider the two other cases where the results are different. These are two endohedral boron atoms in C$_{60}$, i.e. B$_2$@C$_{60}$,
and the structure with one boron atom inside C$_{60}$ and the other boron atom outside it, i.e. B@C$_{60}$B.
For B$_2$@C$_{60}$ the ground state is the structure 2,0(mol) shown in Fig.~4(e,f) of \cite{Bib}, and the configuration Iso1 in lower left panel of
Fig.~1 of the Comment \cite{Xu}.
In Table \ref{tab1}
we list the results of additional HF/6-31G, HF/6-31G* and HF-RIMP2/6-31G calculations at two different geometries: the first [2,0(mol)] is taken from
our study, Ref.~\cite{Bib}, while the second (Iso1) -- from
the supplement coordinate data-file of Ref.~\cite{Xu}. (Here RIMP2 stands for the resolution of identity (RI) version of MP2.)
We see that at the level of HF/6-31G and HF/6-31G* the minima found in our study have {\it lower} energies, which
once again confirms the correctness of our geometry optimization results within the HF method.
The same applies to the B@C$_{60}$B case, Table~\ref{tab2}. In our study the ground state is the configuration (1,1) in Fig.~6 of \cite{Bib},
in the Comment -- Iso1 in the lower middle panel of Fig.~1 of Ref.~\cite{Xu}.
For B$_2$@C$_{60}$ our minimum remains the lowest in RIMP2, while RIMP2 for B@C$_{60}$B indicates that the Iso1 is the lowest.

Xu and Hou also criticize us for using the 6-31G basis set lacking the polarization $d-$functions.
Here it is worth mentioning that we have chosen this relatively poor basis only to be able to handle and complete the MP2 calculations.
As well known, for open electron shells the HF-MP2 method
requiring much memory and the CPU time, is laborious and time consuming.
In the Comment Xu and Hou pointed out that this basis set could lead to qualitative errors.
Although in principle this is conceivable, in the present case such a situation is not expected by us.
The boron being the left neighbor of carbon in the periodic table of elements, experiences the covalent bonding with it
mainly through the valence atomic $2s-$ and $2p-$functions,
which are already included in the 6-31G basis set. The polarization $d-$functions in this binding play only a secondary role.
This viewpoint is supported by our calculations presented in Tables~\ref{tab1} and \ref{tab2}.
The optimal geometries found with HF/6-31G in \cite{Bib} remain the lowest also in HF/6-31G*. 
Additional structural optimization at the HF/6-31G* level yields tiny changes in the geometry and total energy of 
the ground state: -2320.92123 a.u. for B$_2$C$_{60}$ and -2320.88924 a.u. for BC$_{60}$B, compare with $E_1$ in Tables~\ref{tab1} and \ref{tab2}. 

It is worth mentioning that the numerical data quoted in \cite{Xu}, hinder a detailed comparison with our results \cite{Bib}.
From Appendix A of \cite{Xu} it is not clear in which cases restricted and in which unrestricted version of DFT has been employed.
Xu and Hou frequently quote charges of B and C atoms obtained from the natural population analysis.
The effective atomic charge computed from molecular orbitals is known to be an ambiguous characteristic,
and in our work we have used uniquely defined Bader charges \cite{Bader}.

In summary, the results of DFT calculations presented by Xu and Hou in \cite{Xu} should be compared with ours on equal ground,
as done e.g. in the case of benzene dimer \cite{Sher}.
In view of the fact that different variants of DFT result in different ground state geometries, we cannot accept the results
of the Comment as benchmark calculations.
Answering the question on the role of the polarization $d-$functions raised in the Comment,
we have presented here the results of additional calculations using the DFT equilibrium geometries
for B$_2$@C$_{60}$, Table \ref{tab1}, and B@C$_{60}$B, Table \ref{tab2}. (These are two cases where the DFT and HF geometry optimizations
predict different ground state confirmations.)
From Tables \ref{tab1} and \ref{tab2} it follows that
in HF-method DFT-confirmations have higher energies.
Therefore, in Ref.\ \cite{Bib} there is no mistake in finding structures with lowest energies within the chosen method (HF).
The conclusion holds for both 6-31G and 6-31G* basis sets, which shows that the polarization functions
do not change the order of energy minima.

Finally, we remark that the authors of the Comment sometimes use imprecise or incorrect formulations.
For example, as we discussed earlier, the results of DFT and HF-MP2 coincide in three out of five total cases (not in ``couple of cases"),
in Ref.~\cite{Bib} we have not concluded that a single boron atom is located at ``double bond midpoint between two hexagons'',
which is simply wrong, etc.

\end{document}